\documentclass[prd,superscriptaddress,showpacs,nofootinbib,amsmath,amssymb,aps,10pt]{revtex4}

\usepackage{bm}
\usepackage{amsfonts}
\usepackage{latexsym}
\usepackage[latin1]{inputenc}
\usepackage{graphicx}
\usepackage{amsmath}
\usepackage{palatino}
\usepackage{mathpazo}
\usepackage{epstopdf}
\usepackage{booktabs}
\usepackage{dcolumn}
\usepackage{array}

\usepackage{wrapfig}
\usepackage{wasysym}

\def\jnl@style{\it}
\def\aaref@jnl#1{{\jnl@style#1}}

\def\aaref@jnl#1{{\jnl@style#1}}

\def\aj{\aaref@jnl{AJ}}                   
\def\apj{\aaref@jnl{ApJ}}                 
\def\apjl{\aaref@jnl{ApJ}}                
\def\apjs{\aaref@jnl{ApJS}}               
\def\apss{\aaref@jnl{Ap\&SS}}             
\def\aap{\aaref@jnl{A\&A}}                
\def\aapr{\aaref@jnl{A\&A~Rev.}}          
\def\aaps{\aaref@jnl{A\&AS}}              
\def\mnras{\aaref@jnl{Mon.~Not.~Roy.~Astron.~Soc.}}             
\def\prd{\aaref@jnl{Phys.~Rev.~D}}        
\def\prc{\aaref@jnl{Phys.~Rev.~C}}  
\def\prl{\aaref@jnl{Phys.~Rev.~Lett.}}    
\def\qjras{\aaref@jnl{QJRAS}}             
\def\skytel{\aaref@jnl{S\&T}}             
\def\ssr{\aaref@jnl{Space~Sci.~Rev.}}     
\def\zap{\aaref@jnl{ZAp}}                 
\def\nat{\aaref@jnl{Nature}}              
\def\aplett{\aaref@jnl{Astrophys.~Lett.}} 
\def\apspr{\aaref@jnl{Astrophys.~Space~Phys.~Res.}} 
\def\physrep{\aaref@jnl{Phys.~Rep.}}      
\def\physscr{\aaref@jnl{Phys.~Scr}}       
\def\commat{\aaref@jnl{Comm.~Math.~Phys.}}              
\def\science{\aaref@jnl{Science}}               
\def\cqg{\aaref@jnl{Classical Quant.~Grav.}}            
\def\jpcs{\aaref@jnl{JPCS}}                                     
\def\ijmpd{\aaref@jnl{Int.~J.~Mod.~Phys.~D}}                    
\def\grg{\aaref@jnl{Gen.~Relat.~Gravit.}}               
\def\rpp{\aaref@jnl{Rep.~Prog.~Phys.}}          
\def\npa{\aaref@jnl{Nucl.~Phys.~A}}        
\def\lrr{\aaref@jnl{Living Rev.~Rel.}}                   
\def\jcap{\aaref@jnl{J.~Cosmology Astropart.~Phys.}}    
\def\rmp{\aaref@jnl{Rev.~Mod.~Phys.}}   

\allowdisplaybreaks[1]

\begin{document}

\title{Static and slowly rotating neutron stars in scalar-tensor theory with self-interacting massive scalar field}

\author{Kalin V. Staykov}
\email{kstaykov@phys.uni-sofia.bg}
\affiliation{Department of Theoretical Physics, Faculty of Physics, Sofia University, Sofia 1164, Bulgaria}

\author{Dimitar Popchev}
\affiliation{Department of Theoretical Physics, Faculty of Physics, Sofia University, Sofia 1164, Bulgaria}

\author{Daniela D. Doneva}
\email{daniela.doneva@uni-tuebingen.de}
\affiliation{Theoretical Astrophysics, Eberhard Karls University of T\"ubingen, T\"ubingen 72076, Germany}
\affiliation{INRNE - Bulgarian Academy of Sciences, 1784  Sofia, Bulgaria}

\author{Stoytcho S. Yazadjiev}
\email{yazad@phys.uni-sofia.bg}
\affiliation{Department of Theoretical Physics, Faculty of Physics, Sofia University, Sofia 1164, Bulgaria}
\affiliation{Institute of Mathematics and Informatics, Bulgarian Academy of Sciences, Acad. G. Bonchev Street 8, Sofia 1113, Bulgaria}


\begin{abstract}
Binary pulsar observations and gravitational wave detections seriously constrained scalar-tensor theories with massless scalar field allowing only small deviations from general relativity. If we consider a nonzero mass of the scalar field, though, significant deviations from general relativity are allowed for values of the parameters that are in agreement with the  observations. In the present paper we extend this idea and we study scalar-tensor theory with massive field with self-interaction term in the potential. The additional term suppresses the scalar field in the neutron star models in  addition to the effect of the mass of the scalar field but still, large deviations from pure GR can be observed for values of the parameters that are in agreement with the
observations.  
\end{abstract}

\pacs{}
\maketitle
\date{}

\section{Introduction}
Over the last few decades the scalar-tensor theories (STT) of gravity have been studies as the most natural cosmological and astrophysical generalization of general relativity (GR). Particularly interesting subclasses of STT are those for which the weak field regime coincides with GR, and deviations would occur only for strong gravity, i.e. in the gravitation field of compact objects like black holes and neutron stars. For the considered classes of STT, however, black holes are ruled out of the picture due to the existence of "no-hair" theorems, which makes neutron stars the perfect natural laboratory to test strong gravity regime, and modified theories.  Neutron star structure, properties, and physical effects in classes of STT where scalarization of the solutions is observed in the strong field regime, were extensively studied in the past decades (see e.g. \cite{Damour1993,Damour1996,Harada1997,Harada1998,Salgado1998,Pani2011,Sotani2012,Doneva2013,Motahar2017}) both in the static and rapidly rotating cases. A few year ago particular interest attracted the STT with massless scalar field \cite{Popchev2015,Ramazanoglu2016,Yazadjiev2016,Doneva:2016xmf} due to the possibilities for much larger deviations from GR compared to the massless case within the observationally allowed values of the parameters.

The effect of spontaneous scalarization of neutron stars (and if one considers more exotic objects -- quark stars) has nonperturbative scale and thus the deviations from pure general relativity can be very large. In the recent years, however, the astrophysical observations constrained significantly the massless STT \cite{Freire2012,Antoniadis2013}. Thus, the parameters of these theories were seriously restricted to narrow sets which does not allow for significant physical deviations from pure GR.  If one, however, extends the study to the case of STT with massive scalar field, the situation changes dramatically and the parameters in this case are only weakly restricted. The reason lays in the fact that for a scalar field with mass $m_{\varphi}$ one can assign a Compton wave-length $\lambda_{\varphi} = 2\pi/m_{\varphi}$, which leads to a finite range of the scalar field. More precisely the presence of the scalar field will be suppressed outside the compact object at distance greater than the corresponding Compton wave-length of the field $\lambda_{\varphi}$. This means that observations of compact objects with a scale greater than $\lambda_{\varphi}$ can not set any rigorous constraints on the parameters of massive STTs  \cite{Ramazanoglu2016,Yazadjiev2016} (see also \cite{Alsing2012} for the massive Brans-Dicke case).

The most popular class of STT which exhibits the nonperturbative strong field effect of spontaneous scalarization has an Einstein frame coupling function of the form $\alpha(\varphi) = \beta \varphi$, with $\beta$ being negative constant. The observations of binary systems set very rigid constraints on the free parameter in the massless scalar field case, namely $\beta \gtrsim -4.5$ \cite{Demorest10,Freire2012,Antoniadis2013} which is quite restrictive if one considers that spontaneous scalarization can be observed only for $\beta \lesssim -4.35$ for static neutron stars \cite{Damour1992,Damour1996,Harada1998} and  $\beta \lesssim -3.9$ for rapidly rotating stars \cite{Doneva2013}. The allowed values for the parameter $\beta$ can be much smaller than $-4.5$ for massive STT and detailed considerations of the problem are given in \cite{Ramazanoglu2016,Yazadjiev2016}. For example, the strongest constraint on massless STT comes from the observations of the pulsar-white dwarf binary ${PSR \, J0348+0432}$ \cite{Antoniadis2013} and if we consider a massive scalar field with  $m_{\varphi} \gg 10^{-16} eV$, which is equivalent to Compton wave-length $\lambda_{\varphi} \ll 10^{10} m$, then ${PSR \, J0348+0432}$ practically can not impose any constraints on the parameter $\beta$ since $\lambda_{\varphi}$ is smaller than the orbital separation between the two stars. In addition the requirement that we can have scalarized neutron star  but no scalarization for white dwarfs leads to  $3 \lesssim -\beta \lesssim 10^{3}$ \cite{Ramazanoglu2016}. 

As one can see in the presence of massive scalar field the observationally allowed values for $\beta$ can significantly differ from the massless STT ones, and neutron stars in both cases can have significantly different properties and structure. This was thoroughly investigated  for static and for slowly rotating neutron stars in \cite{Popchev2015,Ramazanoglu2016,Yazadjiev2016,Doneva:2016xmf,Motahar2017}. A natural extension in this case is to include self-interaction of the massive scalar field and investigate the influence on the neutron star structure and properties, that would be the goal of the present paper.

The structure of the paper is as follow. In section II we present the basic equations for massive STT, and we introduce the explicit form of the potential for the specific theory we study. In section III the numerical results with some additional comments on the parameters of the theory are presented. The paper ends with a Conclusion. 

\section{Basic equations}

For mathematical simplicity in this paper we work not in the physical Jordan frame, but in the more convenient Einstein frame. All of the results in the next section, however, are presented in the physical Jordan frame. 

The scalar-tensor theory action in the Einstein frame is  given by 

\begin{eqnarray}\label{EFA}
S=\frac{1}{16\pi G} \int d^4x \sqrt{-g}\left[ R - 2
g^{\mu\nu}\partial_{\mu}\varphi \partial_{\nu}\varphi - V(\varphi)
\right] + S_{\rm
	matter}(A^2(\varphi)g_{\mu\nu},\chi),
\end{eqnarray}
where $R$ is the Ricci scalar curvature with respect to the metric $g_{\mu \nu}$. The STTs are specified by the function $A(\varphi)$ and the potential   $V(\varphi)$. In the present paper we shall restrict our study to STT with 
\begin{equation}
A(\varphi)=e^{\frac{1}{2}\beta \varphi^2}
\end{equation} 
and a non-negative scalar potential with self-interaction. The natural and the simplest case is a $Z_2$ symmetric 
scalar potential with a quartic self-interaction, namely    
\begin{equation}
V(\varphi)=2m^2_{\varphi}\varphi^2 + \lambda \varphi^4,
\end{equation}
where $m_{\varphi}$  is the mass of the scalar field $\varphi$ and $\lambda\ge 0$ is a parameter with dimension of $length^{-2}$. This particular choice of $A(\varphi)$ leads to a STT that is indistinguishable from pure GR in the weak field regime, while non-perturbative effects can appear for strong fields. The first term in the scalar field potential $V(\varphi)$ on the other hand is the standard one considered in previous studies of massive STT \cite{Popchev2015,Ramazanoglu2016,Yazadjiev2016,Doneva:2016xmf} while the second term describes self-interaction of the scalar field and was not considered until now. 

The Jordan frame metric ${\tilde g}_{\mu\nu}$ is connected to the Einstein one $g_{\mu\nu}$ via the conformal relation ${\tilde g}_{\mu\nu}= A^2(\varphi)g_{\mu\nu}$  and the gravitational scalar respectively by $\Phi=A^{-2}(\varphi)$. The relation between the energy-momentum tensor in both frames is given by the formula $T_{\mu\nu}=A^2(\varphi){\tilde T}_{\mu\nu}$, where  $T_{\mu\nu}$ is the Einstein frame one, and  ${\tilde T}_{\mu\nu}$ is the Jordan frame one.
In the case of a perfect fluid the relations between the energy density and pressure in both frames are given by $\rho=A^4(\varphi){\tilde \rho}$ and $p=A^4(\varphi){\tilde p}$.

In this paper we are using slow rotation approximation in first order in the angular velocity $\Omega$, i.e. keeping only first order terms. Furthermore we consider stationary and axisymmetric spacetime as well as stationary and axisymmetric scalar field and  fluid configurations. The spacetime metric in this case is taken in the standard form \cite{Hartle1967}

\begin{eqnarray}
ds^2= - e^{2\phi(r)}dt^2 + e^{2\Lambda(r)}dr^2 + r^2(d\theta^2 +
\sin^2\theta d\vartheta^2 ) - 2\omega(r,\theta)r^2 \sin^2\theta  d\vartheta dt.
\end{eqnarray}

While the metric function $\omega$ is in linear order of $\Omega$, the rotational corrections to other metric functions, the scalar field,
the fluid energy density and pressure are of order ${\cal O}(\Omega^2)$. That is why within this approximation we can derive the moment of inertia of the star, but the rest of the parameters, such as the mass and the  radius, coincide with the ones of a static model. 

The dimensionally reduced Einstein frame field equations, derived from the action (\ref{EFA}) and containing at most terms linear in $\Omega$,  are the following

\begin{eqnarray} \label{eq:FieldEq}
&&\frac{1}{r^2}\frac{d}{dr}\left[r(1- e^{-2\Lambda})\right]= 8\pi G
A^4(\varphi) {\tilde \rho} + e^{-2\Lambda}\left(\frac{d\varphi}{dr}\right)^2
+ \frac{1}{2} V(\varphi), \nonumber\\
&&\frac{2}{r}e^{-2\Lambda} \frac{d\phi}{dr} - \frac{1}{r^2}(1-
e^{-2\Lambda})= 8\pi G A^4(\varphi) {\tilde p} +
e^{-2\Lambda}\left(\frac{d\varphi}{dr}\right)^2 - \frac{1}{2}
V(\varphi),\nonumber\\
&&\frac{d^2\varphi}{dr^2} + \left(\frac{d\phi}{dr} -
\frac{d\Lambda}{dr} + \frac{2}{r} \right)\frac{d\varphi}{dr}= 4\pi G
\alpha(\varphi)A^4(\varphi)({\tilde \rho}-3{\tilde p})e^{2\Lambda} + \frac{1}{4}
\frac{dV(\varphi)}{d\varphi} e^{2\Lambda}, \\
&&\frac{d{\tilde p}}{dr}= - ({\tilde \rho} + {\tilde p}) \left(\frac{d\phi}{dr} +
\alpha(\varphi)\frac{d\varphi}{dr} \right), \nonumber\\
&&\frac{e^{\Phi-\Lambda}}{r^4} \partial_{r}\left[e^{-(\Phi + \Lambda)} r^4 \partial_{r}{\bar\omega} \right]  + \frac{1}{r^2\sin^3\theta} \partial_{\theta}\left[\sin^3\theta\partial_{\theta}\bar\omega \right]= 16\pi GA^4(\varphi)({\tilde \rho} + {\tilde p})\bar\omega , \nonumber
\end{eqnarray}
where the function $\bar\omega$ is defined as $\bar\omega = \Omega - \omega$, and the coupling function $\alpha(\varphi)$ is defined by $\alpha(\varphi)=\frac{d\ln A(\varphi)}{d\varphi}$.

The system of equations (\ref{eq:FieldEq}),  supplemented with the equation of state for the matter inside the star and the appropriate boundary conditions, describes the interior  and the exterior of the neutron star. For the exterior of the neutron star to be described by the system (\ref{eq:FieldEq}), however, we have to set ${\tilde \rho}={\tilde p}=0$.

The natural boundary conditions at the center of the star are $\rho(0)=\rho_{c},  \Lambda(0)=0, {\rm and} \frac{d{\varphi}}{dr}(0)= 0$, where $\rho_{c}$ is the constant central density, while from the requirement for asymptotic flatness, at infinity we have $\lim_{r\to \infty}\phi(r)=0, \lim_{r\to \infty}\varphi (r)=0$ (see e.g. \cite{Yazadjiev2014}). The coordinate radius $r_S$ of the star in the Einstein frame is determined by the standard condition $p(r_S)=0$, while  the physical radius of the star in the Jordan frame is given by $R_{S}= A[\varphi(r_S)] r_S$.

The equation for $\bar \omega$ is separated from the other equations in the system (\ref{eq:FieldEq}) and it can be considerably simplified. Expanding $\bar \omega$ in the form \cite{Hartle1967}

\begin{eqnarray}
\bar\omega= \sum^{\infty}_{l=1}{\bar \omega}_{l}(r) \left(- \frac{1}{\sin\theta}\frac{dP_{l}}{d\theta} \right),
\end{eqnarray}
where $P_{l}$ are Legendre polynomials and substituting into the equation for $\bar \omega$ we find

\begin{eqnarray}\label{OL}
\frac{e^{\Phi-\Lambda}}{r^4} \frac{d}{dr}\left[e^{-(\Phi+ \Lambda)}r^4 \frac{d{\bar\omega}_{l}(r)}{dr} \right] - \frac{l(l+1)-2}{r^2} {\bar\omega}_{l}(r)=
16\pi G A^4(\varphi)(\rho + p){\bar\omega}_{l}(r).
\end{eqnarray}

One can easily show that the asymptotic behavior of the function $\omega$ at large distances from the center of the star and for asymptotically flat spacetimes, has the form ${\bar \omega}_{l} \to {\rm const}_1\, r^{-l-2} + {\rm const}_2\, r^{l-1}$. This asymptotic is also connected with the angular momentum of the star $J$ via the standard relation $\omega \to 2J/r^3$ (or equivalently $\bar\omega \to \Omega - 2J/r^3 $) for $r\to \infty$. Comparing the two expressions for $\omega$ , we conclude that $l=1$, i.e. ${\bar \omega}_{l}=0$ for $l\ge 2$. Therefore, $\bar\omega$ is a function of $r$ only and the equation for $\bar \omega$  is
\begin{eqnarray}\label{OR}
\frac{e^{\Phi-\Lambda}}{r^4} \frac{d}{dr}\left[e^{-(\Phi+ \Lambda)}r^4 \frac{d{\bar\omega}(r)}{dr} \right] =
16\pi G A^4(\varphi)(\rho + p){\bar\omega}(r).
\end{eqnarray}
The natural boundary condition for ${\bar\omega}$ to ensure its regularity at the center of the star is $\frac{d{\bar\omega}}{dr}(0)= 0$, and at infinity $\lim_{r\to \infty}{\bar\omega}=\Omega$. 

As we mentioned earlier, in the present paper we consider the moment of inertia $I$ of the compact star. It is defined in the standard way 

\begin{eqnarray}
I=\frac{J}{\Omega}.
\end{eqnarray}
Using  equation (\ref{OR}) for $\bar \omega$  and the asymptotic form of $\bar \omega$  one can also derive a more convenient for numerical computations integral equation for the moment of inertia 

\begin{eqnarray}\label{eq:I_integral}
I= \frac{8\pi G}{3} \int_{0}^{r_S}A^4(\varphi)(\rho + p)e^{\Lambda - \Phi} r^4 \left(\frac{\bar\omega}{\Omega}\right) dr .
\end{eqnarray}

In the next section where we present our numerical results we shall use the dimensionless parameters $m_{\varphi}\to m_{\varphi} R_{0}$ and $\lambda \to \lambda R_{0}^2$, where $M_{\odot}$ is the solar mass and $R_{0}=1.47664 \,{\rm km}$ is one half of the solar gravitational radius.

\section{Numerical results}

As we have mentioned earlier, binary systems of compact objects \cite{Freire2012,Antoniadis2013} are used to set rigorous constraints on the parameters in the massless STT with $A(\varphi) = e^{\beta\varphi^2/2}$, leaving the possibility only for small deviations from GR by setting $\beta > -4.5$, while spontaneous scalarization occur for $\beta < - 4.35$ \cite{Damour1996,Harada1998} and $\beta < -3.9$ \cite{Doneva2013} for static and for rapidly rotating models correspondingly. This, however, significantly changes if massive scalar field is added. The constraints of these theories come from observations on shrinking of the orbit of the binaries due to gravitational wave emission and the theory free parameters should be in agreement with these observations. More precisely, the emitted gravitational radiation match very well the GR predictions, which means that there is non or negligible scalar gravitational  radiation. As a result the observed objects should be nonscalalrized or very weakly scalarized. If one considers massive scalar field, however, the mass of the scalar field suppresses the emission of scalar radiation, which reconciles already discarded values of $\beta$ with the binary observations. The lower boundary for the mass of the scalar field can be set by these same binary systems \cite{Freire2012,Antoniadis2013}, and more precisely by the distance between the two companions ($r_{binary}$). In order to have negligible scalar gravitational radiation, the Compton wave-length of the field should be smaller than the orbital separation $\lambda_{\varphi} \ll r_{binary}$. For the observed binaries $r_{binary} \sim 10^9 \rm{m}$, which translates into $m_{\varphi} \gg 10^{-16} \rm{eV}$. The upper limit for the mass of the field should be such that it does not suppresses the spontaneous scalarization in the stars, i.e. the characteristic length of the star should be smaller than the Compton wave-length for the corresponding field. In numbers this translates as $m_{\varphi} \lesssim 10^{-9} {\rm{eV}}$. As a final interval for the mass $m_{\varphi}$ of  the scalar field we have

\begin{equation}
10^{-16} {\rm eV} \lesssim m_{\varphi} \lesssim 10^{-9} {\rm eV},
\end{equation}  
which roughly corresponds to $10^{-6} \lesssim m_{\varphi} \lesssim 10$ in our dimensionless units.
Although, there are additional midrange constraints for the mass of the scalar field, the above ones are the most reliable and we will stick to them. 

We already have mentioned that if the mass of the scalar field is sufficiently large, the parameter $\beta$ can be set in significantly wide interval of values compared to the massless case, more precisely $3 \lesssim -\beta \lesssim 10^{3}$ coming from the requirement that we can have scalarized neutron stars but no scalarization for white dwarfs. We will, however, consider only moderate values of $\beta \geq -10$, on one side to be in correlation with \cite{Ramazanoglu2016,Popchev2015,Yazadjiev2016}, and on the other we have additional parameter $\lambda$ coming from the self-interaction term in the potential, and it is a good practice to study its effect for familiar and well behaving models. Concerning the parameter $\lambda$ we  constrain ourself to values which allow spontaneous scalarization.

In this paper we employ one of the most popular EOS, the so-called APR4 EOS \cite{AkmalPR}, for which the piecewise polytropic approximation is used \cite{Read2009}. We will concentrate on the manifestation of the free parameters in the theory instead of considering a wide variety of EOS since here we have a three parameter ($\beta$, $m_\varphi$ and $\lambda$) family of solutions. The neutron star models are studied in slow rotations approximation in first order in $\Omega$, which means that we can determine the moment of inertia of the star, but the mass and the radius does not change with respect to the static case because the corrections to these quantities are of second order in $\Omega$. The system of equations (\ref{eq:FieldEq}) combined with the EOS is solved using a shooting method, where the central value of the scalar field $\varphi$, and the metric functions $\phi$ and $\omega$ are the shooting parameters. 

\begin{figure}[]
	\centering
	\includegraphics[width=0.45\textwidth]{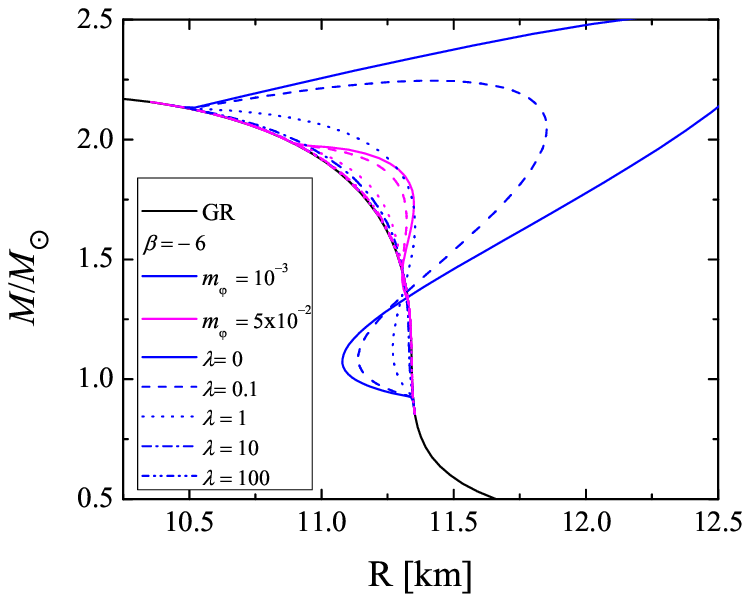}
	\includegraphics[width=0.45\textwidth]{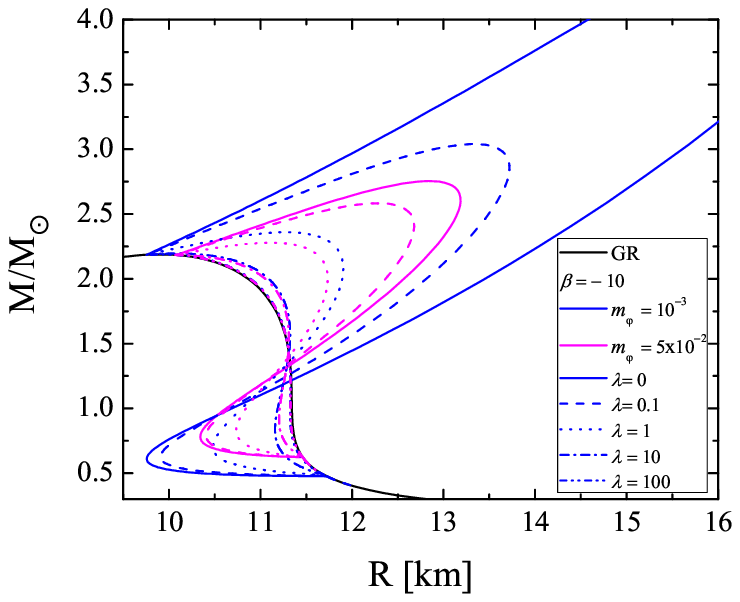}
	\caption{Mass of radius relation for models with EOS APR4, $\beta = - 6$ (left), and $\beta = -10$ (right). On both figures are presented results for GR (black continuous line), mass of the scalar field $m_{\varphi} = 10^{-3}$ with $\lambda = 0$ (blue continuous line) and $m_{\varphi} = 10^{-3}$ with different values for $\lambda$ (blue lines in different patterns), and with mass of the scalar field $m_{\varphi} = 5\times 10^{-2} \,\,{\rm{with}}\,\, \lambda = 0$ (purple continuous line) and $m_{\varphi} = 5\times 10^{-2}$ with different values for $\lambda$ (purple lines in different patterns).  }
	\label{Fig:M_R}
\end{figure}

\begin{figure}[]
	\centering
	\includegraphics[width=0.45\textwidth]{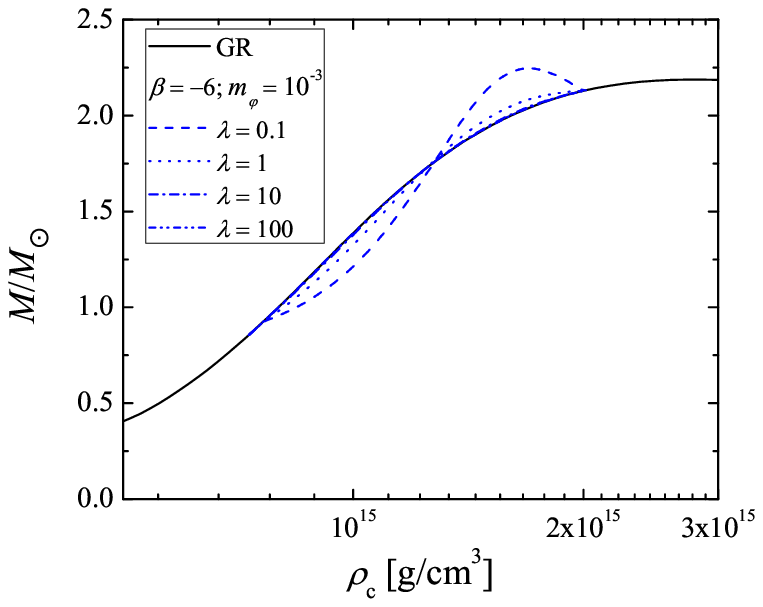}
	\includegraphics[width=0.45\textwidth]{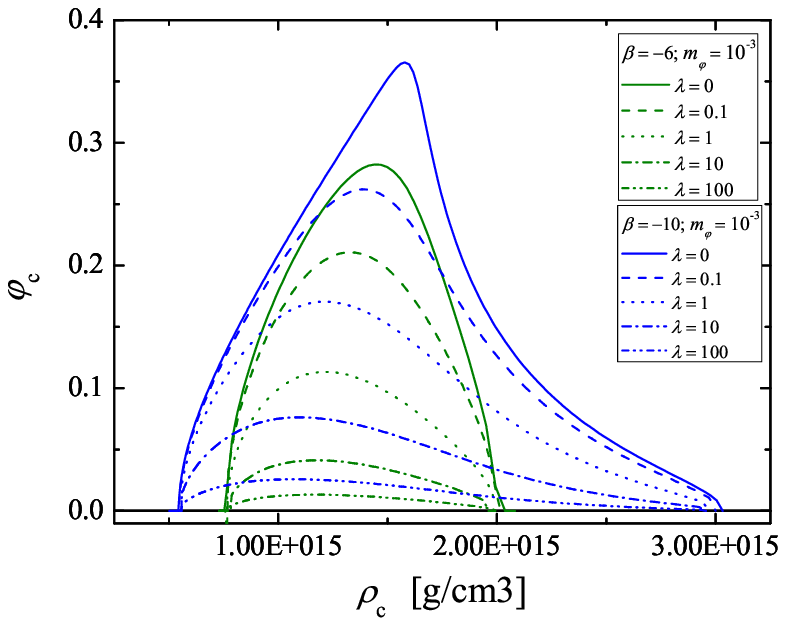}
	\caption{(\textit{Left}) Mass as function of the central density of the models. The results are for GR (black), and $\beta = -6$ with $m_{\varphi} = 10^{-3}$ and different values for $\lambda$ (blue in different patterns). (\textit{Right}) The central value of the scalar field as a function of the central density of the model.  The models are with $\beta = -6$ (blue) and $\beta = -10$ (dark green), $m_{\varphi} = 10^{-3}$ and different values for $\lambda$ (in different patterns).}
	\label{Fig:M_rho}
\end{figure}

\begin{figure}[]
	\centering
	\includegraphics[width=0.45\textwidth]{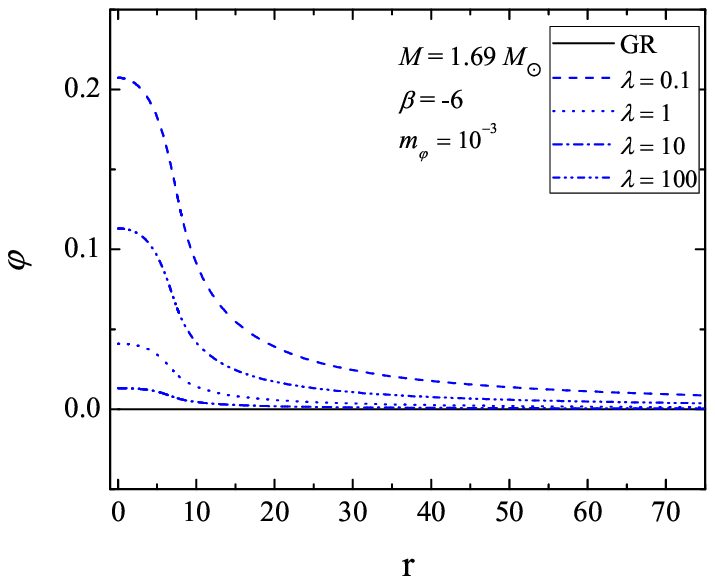}
	\includegraphics[width=0.45\textwidth]{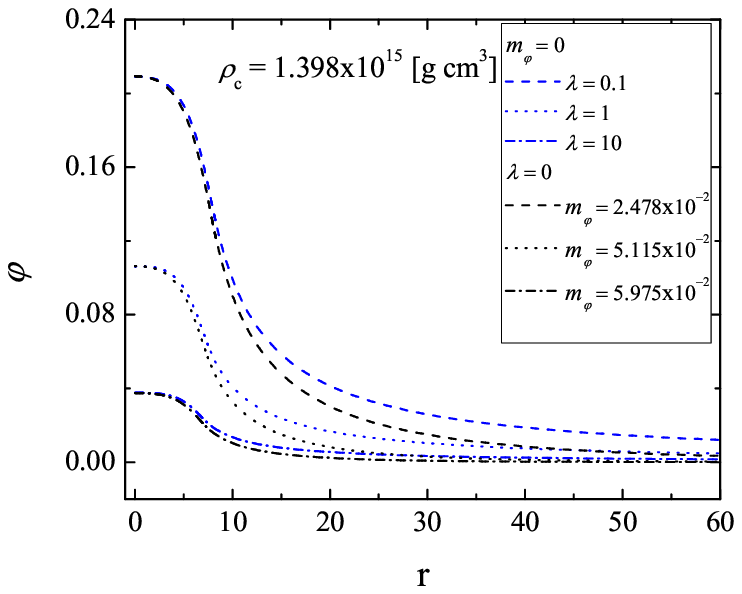}
	\caption{(\textit{Left}) The scalar field distribution with the radial coordinate for models with equal mass.  The models are with $\beta = -6$ and $m_{\varphi} =  10^{-3}$ and different values for $\lambda$ (blue in different patterns). (\textit{Right}) The radial distribution of the scalar field for models with equal central density. The results are for massless STT with self-interaction (blue), and for massive STT without self-interaction (black).}
	\label{Fig:phi_r}
\end{figure}

In Fig. \ref{Fig:M_R} we plot the mass of radius relation for two values for the parameter $\beta$ (the left and the right panel). Different combinations of the mass of the scalar field and the value of the coupling constant $\lambda$ (not to be mistaken with the Compton wave-length of the scalar field) are presented. In both panels some of the results for the pure massive case, i.e. $\lambda = 0$ (continuous blue lines) are partially cut out of the figures in order to have a better visibility of the results for different nonzero $\lambda$. As one can see the self-interaction term in the potential additionally suppresses the scalarization in the star. For all of the rest parameters fixed, the limiting case of $\lambda \rightarrow 0$ leads to the results for massive STT with quadratic term (respectively to the massless case if the massive term is absent), and with the increases of $\lambda$ the results converge to the GR ones. One can examine this behavior further in Fig. \ref{Fig:M_rho}, where the mass as function of the central density (left) and the central value for the scalar field as function of the central density (right) are plotted. It is interesting to point out that with the increase of the parameter $\beta$, the shape of relation $\varphi_c(\rho_c)$ changes and a sharp maximum of $\varphi_c$ can be observed. In Fig. \ref{Fig:phi_r} we plot the distribution of the scalar field with the radial coordinate. In the left panel we study the scalar field for fixed value of the mass of the field, fixed $\beta$, and fixed mass of the models with different values for $\lambda$. In the right one we study models with equal central density for the massless case with different values for $\lambda$, and models without self-interaction with different values for the mass of the field. The models in the last figure are pared two by two for equal central values of the scalar field in order to examine the effect of the different terms in the potential individually.  The expected  decay of the scalar field is clearly visible in both panels but it is clear that for massless field with self-interaction the scalar field decay is slower. In Table \ref{Tbl:1} we present the parameters of these models. It is clear that for the same central density and scalar field, the self-interaction term has marginally more pronounced effect on the mass of the star and its radius, which considering the distribution of the scalar field is natural (the more slowly decaying scalar field will have higher contribution to the gravitational mass of the star).  

\begin{table}[]
	\begin{tabular}{lccc}
		\quad & $M/M_{\odot}$ & $R$ [km] & $\varphi_c$ \\
		\hline
		\hline
		GR     & 1.867 & 11.06& 0.0\\
		\hline
		$\lambda = 0.1$, $m_{\varphi} = 0$    & 1.996 & 12.43 & 0.209  \\
		$\lambda = 0$,\;\;\; $m_{\varphi} = 2.478\times 10^{-2}$ & 1.958 & 12.32 & 0.209 \\
		\hline
		$\lambda = 1$,\;\;\; $m_{\varphi} = 0$    & 1.903 & 11.39 & 0.106  \\
		$\lambda = 0$,\;\;\; $m_{\varphi} = 5.115\times 10^{-2}$ & 1.888 & 11.35 & 0.106\\
		\hline
		$\lambda = 10$,\; $m_{\varphi} = 0$ & 1.872 & 11.10 & 0.0376 \\
		$\lambda = 0$, \;\; $m_{\varphi} = 5.975\times 10^{-2}$ & 1.870 & 11.10 & 0.0376\\
		
	\end{tabular}
	\caption{Parameters of the star for GR, massive STT without self-interaction, and massless STT with self interaction for the same central density $\rho_c=1.398\times 10^{15} {\rm [g/cm^3]}$.}
	\label{Tbl:1}
\end{table}

\begin{figure}[]
	\centering
	\includegraphics[width=0.45\textwidth]{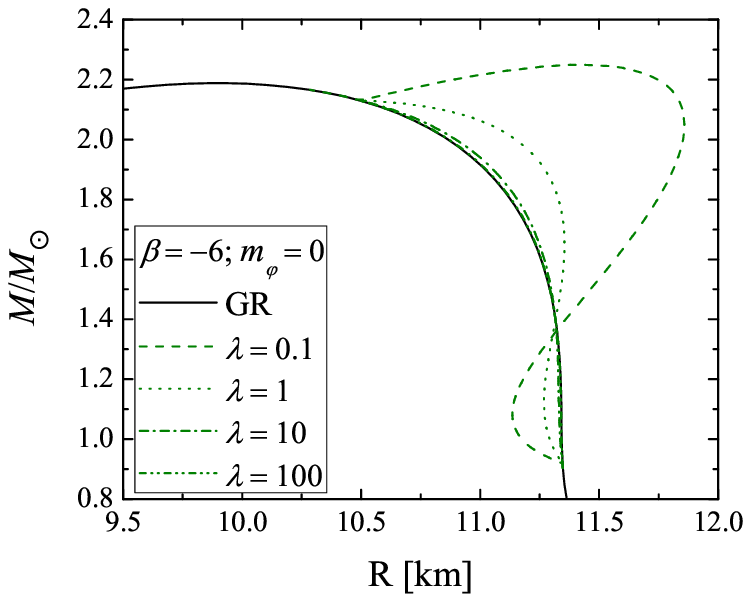}
	\includegraphics[width=0.45\textwidth]{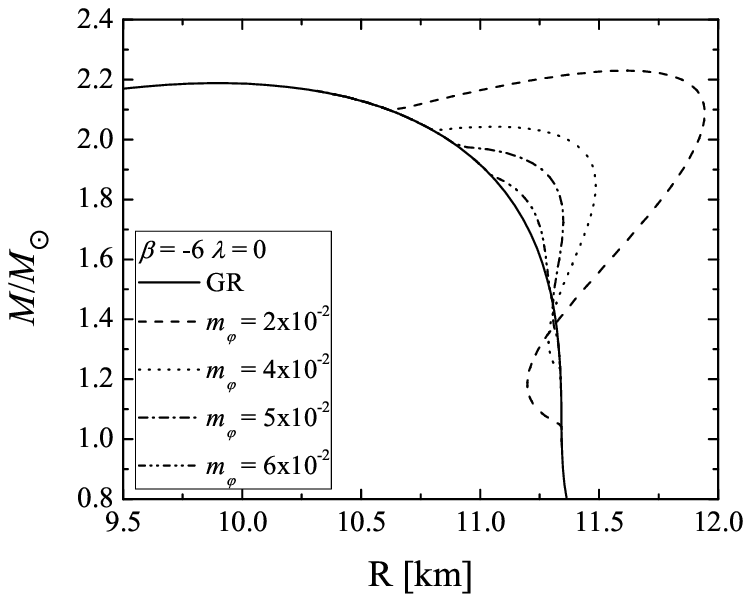}
	\caption{ Mass of radius relation for massless STT with self-interaction (\textit{Left}), and for massive STT without self-interaction (\textit{Right}). Both relations are for $\beta = -6$, and different values of the free parameter.}
	\label{Fig:lvsm}
\end{figure}

In Fig. \ref{Fig:lvsm} we plot the mass of radius relation in two different cases: massless STT with self-interaction (left) and massive STT without self-interaction (right). One can see the expected  consequences from the self-interaction term. Both terms independently suppress scalarizaton, but except for this, the effect is qualitatively different. While the massive term ($\sim \varphi^2$) suppresses the field, in the same time both bifurcation points (the one at lower and the one at higher central energy densities) move to each other. The self interaction term ($\sim \varphi^4$) on the other hand also suppresses the scalar field but it does not change the position of the bifurcation points. The latter means that even for big values for $\lambda$, i.e. highly suppressed scalalrization, we will have wider range of central density values for which scalarization can occur contrary to the massive case.

\begin{figure}[]
	\centering
	\includegraphics[width=0.45\textwidth]{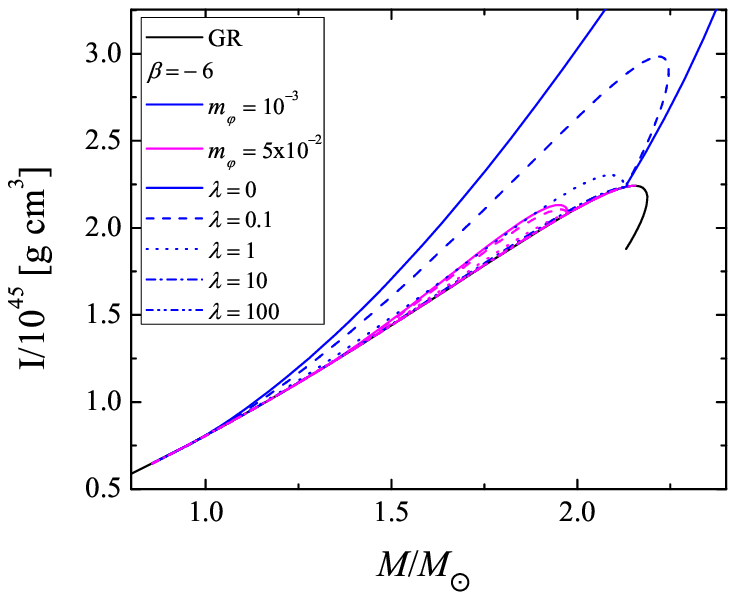}
	\includegraphics[width=0.45\textwidth]{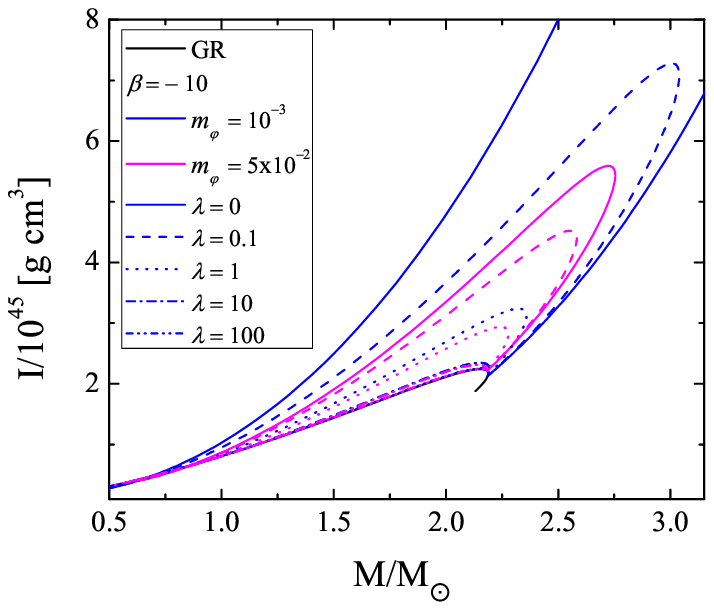}
	\caption{Moment of inertia as function of mass for models with EOS APR4, $\beta = - 6$ (left), and $\beta = -10$ (right). On both figures are presented results for GR (black continuous line), mass of the scalar field $m_{\varphi} = 10^{-3}$ with $\lambda = 0$ (blue continuous line) and $m_{\varphi} = 10^{-3}$ with different values for $\lambda$ (blue lines in different patterns), and with mass of the scalar field $m_{\varphi} = 5\times 10^{-2} \,\,{\rm{with}}\,\, \lambda = 0$ (purple continuous line) and $m_{\varphi} = 10^{-3}$ with different values for $\lambda$ (purple lines in different patterns). }
	\label{Fig:I_M}
\end{figure}

In Fig. \ref{Fig:I_M} we plot the moment of inertia as function of stellar mass for two values for the parameter $\beta$ (the left and the right panel). Different combinations for the mass of the scalar field and for the value of the constant $\lambda$ are shown. As one can see, the self-interaction  term additionally suppresses the scalarization compared to the pure massive case similar to the mass of radius relation presented in Fig. \ref{Fig:M_R}.

Lets us comment on the chosen values for the parameters, and the effect which varying them has on the neutron star models. The chosen values for $\beta$ are smaller compared to the restricted values for the massless STT but they are still quite conservative as compared with the interval of allowed values for STT with massive scalar field. In addition, because of the additionally suppressed scalarization by the self-interaction term, the intervals of allowed values the scalar-field mass can get even wider.  As one can see, the effect of the self-interaction causes partial overlapping of the results for models with low mass of the scalar field and high values for $\lambda$  with models with high mass of the scalar field and low values for $\lambda$, which introduces additional degeneracy between the parameters. In addition, we have examined models with zero mass  of the scalar field, i.e. massless scalar field, but nonzero $\lambda$  and found the behavior to be similar to the case in which we have massive scalar field, and no self-interaction term, but together with the differences discussed above.

\section{Conclusion}

In this paper we studied a certain class of scalar-tensor theories with a massive scalar field with quartic self-interaction term in the potential. The STTs are a natural generalization of Einstein's theory of gravity and they does not suffer from intrinsic problems. The most extensively studied STTs for the last few decades were theories with massless scalar field, but observations of binary systems of compact objects and the gravitational waves emissions drastically restricted the allowed values for the parameters of the theories, which manifests in small deviation from pure GR. 

Adding massive scalar field changes this by reconciling the theory with the observations  for a much wider range for the parameters compared to the massless case. This has been examined in \cite{Popchev2015,Ramazanoglu2016,Yazadjiev2016,Doneva:2016xmf}, and in our paper we extended these studies by including a quartic self-interacting term in the scalar field potential. Our results show that the self-interaction term  additionally suppresses the scalarization, which means it decreases the deviations from GR even more. More precisely, for  fixed value of the couping constant $\beta$ and fixed mass for the scalar field, the deviation from GR decreases if one increases the value for the constant $\lambda$ in the self-interaction term.  This can reconcile  even wider range of values for the scalar-field  masses with the observations. 

In order to study better the effect of the self-interaction term we examined the case of massless scalar field with non-zero self-interaction. The results showed that the scalarization is again  suppressed and up to a large extent the constant in the self-interaction term has qualitatively very similar effect on the neutron star properties as the scalar field mass. The main qualitative difference comes from the fact that the self-interaction does not change the position of the bifurcation points (in the massless case) while the mass of the scalar field  changes the critical values of the parameters where new branches of scalarized solutions appear or disappear.

A standard problem of the alternative theories in general is that modifications of the gravitational theory may either have negligible effect on the neutron star properties or this effect is very similar to the one created by the uncertainty in the equation of state for the matter in the star. In the theory we studied the deviations from pure GR can be considerably larger than the equation of state uncertainty, but addition problem appears. Namely, we have three free parameters of the theory and varying them have very similar effect on the neutron star structure. Thus, breaking the degeneracy between these parameters can not be done solely by the electromagnetic observations of the neutron star mass, radius or moment of inertia. The gravitational wave observations of merging neutron stars, though, might offer additional ways of breaking the degeneracy but further studies in this direction are needed.

\section*{Acknowledgements}

KS, SY, and DD would like to thank for support by the COST Actions CA16214, CA16104 and CA15117.  KS is supported by the Bulgarian NSF Grant DM 18/4. DD would like to thank the European Social Fund, the Ministry of Science, Research and the Arts Baden-W\"{u}rttemberg for the support. DD is indebted to the Baden-W\"{u}rttemberg Stiftung for the financial support of this research project by the Eliteprogramme for Postdocs. DP and SY are supported partially by the Sofia University Grants No 80-10-73/2018 and No 3258/2017.


\bibliography{references}

\end{document}